\documentclass[english,prl,aps,twocolumn,floatfix]{revtex4}

\usepackage[linktocpage,bookmarksopen,bookmarksnumbered]{hyperref}
\usepackage{graphicx}
\usepackage{dcolumn}
\usepackage{amsmath,graphics,epsfig,color,verbatim,ulem}
\usepackage{amssymb}

\usepackage{babel}

\begin{document}

\title{The magnetic excitation spectra in BaFe$_{2}$As$_{2}$: a two-particle
approach within DFT+DMFT}

\author{Hyowon Park, Kristjan Haule, and Gabriel Kotliar}

\affiliation{Department of Physics, Rutgers University, Piscataway, NJ 08854,
USA}

\date{\today}

\begin{abstract}
We study the magnetic excitation spectra in the paramagnetic state of
BaFe$_{2}$As$_{2}$ from the \textit{ab initio}
perspective.
The one-particle excitation spectrum is determined within the
combination of the density functional theory and the dynamical
mean-field theory method.
The two-particle response function is
extracted from the local two-particle vertex function, also computed
by the dynamical mean field theory, and the polarization function.
This method reproduces all the experimentally observed features in
inelastic neutron scattering (INS), and relates them to both the one
particle excitations and the collective modes.
%
%At low frequency the magnetic excitation spectra, as encoded in
%S($q,\omega$), is strongly peaked at the magenetic ordering wave
%vector $(1,0,1)$.  With increasing energy the peak position shifts,
%and around $230\,$meV the peak moves to the wave vector $(1,1,1)$, in
%good agreement with inelastic neutron scattering (INS) experiments on
%BaFe$_2$As$_2$.  This high energy peak is shown to originate mainly
%from the intra-orbital excitations in the $d_{xy}$ orbital from the
%momentum space regions near the two electron pockets.
%
The magnetic excitation dispersion is well accounted for by our theoretical
calculation in the paramagnetic state without any broken symmetry,
hence nematic order is not needed to explain the INS experimental
data.
\end{abstract}

\maketitle

Neutron scattering experiments provide strong constraints on the theory
of iron pnictides. Both the localized picture and the itinerant picture
of the magnetic response have had some successes in accounting or
even predicting aspects of the experiments. Calculations based on
a spin model with frustrated exchange constants~\cite{Goswami:10,Singh:10}
or with biquadratic interactions~\cite{Antropov:11} described well
the neutron scattering experiments~\cite{Diallo:09,Zhao:10}. The itinerant
magnetic model, based on an random phase approximation (RPA) form
of the magnetic response, uses polarization functions extracted from
density functional theory (DFT)~\cite{Ke:11} or tight binding fits~\cite{Keimer:10,Graser:10,Tohyama:10}
and produces equally good descriptions of the experimental data. 

Furthermore, DFT calculations predicted the stripe nature of the
ordering pattern~\cite{Dong:08} and the anisotropic values of the
exchange constants which fit well the spin wave dispersion in the
magnetic phase~\cite{Han:09}. The tight
binding calculations based on DFT bands also predicted the existence of
a resonance mode in the superconducting state~\cite{Maier:09}.

In spite of these successes, both itinerant and localized models
require significant extensions to fully describe the experimental
results.  DFT fails to predict the observed ordered
moment~\cite{Han:09}. Furthermore, adjusting parameters such as the
arsenic height to reproduce the ordered moment, 
leads to a peak in the density of states at the Fermi
level~\cite{Ke:11}, instead of the pseudogap, which is observed experimentally.
%and more significantly the spin spectral weight computed within the
%random phase approximation (RPA) falls well below the experimental
%observations.
The localized picture cannot describe 
the magnetic order  in the FeTe material without introducing
additional longer range exchange constants. Given that this material
is more localized than the 122, the exchange constants would be
expected to be shorter range. Furthermore, fits of the INS data
require the use of anisotropic exchange constants  well above the magnetic
ordering temperature~\cite{Dai:10}. However no clear phase transition
to a nematic phase in this range has been detected.

In this Letter, we argue that the combination of density functional
theory and dynamical mean field theory (DFT+DMFT) provides a natural
way to improve both the localized and the itinerant picture, and
connects the neutron response to structural material specific
information
and to the results of other spectroscopies. 

\begin{figure}
\includegraphics[scale=0.17]{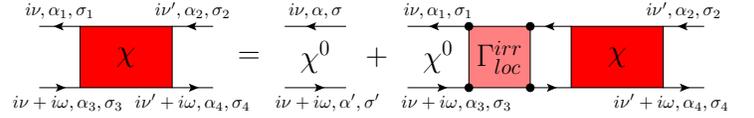}
\caption{(Color online) The Feynman diagrams for the Bethe-Salpeter equation.
  It relates the two-particle Green's function ($\chi$) with the
  polarization ($\chi^{0}$) and the local irreducible vertex function
  ($\Gamma_{loc}^{irr}$).   The non-local two-particle Green's
  function is obtained by replacing
  the local propagator by the non-local propagator.
  \label{fig:BSE}}
\end{figure}

We compute the one-particle Green's function using the charge
self-consistent full potential DFT+DMFT method, as implemented in
Ref.~\cite{Haule:10}, based on Wien2k code~\cite{Wien2K}. We used
the continuous-time quantum Monte Carlo (CTQMC)~\cite{Haule:07,Werner:06} as
the quantum impurity solver, and the Coulomb interaction
matrix as determined in Ref.~\onlinecite{Kutepov:10}.
The dynamical magnetic susceptibility
$\chi(\textbf{q},\omega)$ is computed from the \textit{ab initio} perspective
by extracting the two-particle vertex functions of DFT+DMFT
solution $\Gamma_{loc}^{irr}$~\cite{Jarrell:92}.
% The polarization function $\chi^{0}(q,\omega)$ encodes the
% particle-hole excitations of quasi-particles, and also includes the
% incoherent contribution from the localized moments, as both are
% present in the DMFT solution.
The polarization bubble $\chi^{0}$ is computed from the fully
interacting
one particle Greens function. 
The full susceptibility is computed from  $\chi^{0}$ and  
the two-particle  irreducible vertex function
$\Gamma_{loc}^{irr}$, which is assumed to be local in the same
basis  in which the DMFT self-energy is local, implemented here by the
projector to the muffin-thin sphere~\cite{Haule:10}.  In order to
extract $\Gamma_{loc}^{irr}$, we employ the Bethe-Salpeter equation
(see Fig.~\ref{fig:BSE}) which relates the local two-particle Green's
function ($\chi_{loc}$), sampled by CTQMC, with both the local
polarization function ($\chi_{loc}^{0}$) and $\Gamma_{loc}^{irr}$:
\begin{equation}
\Gamma_{loc{\alpha_{1}\sigma_{1},\alpha_{2}\sigma_{2}\atop
    \alpha_{3}\sigma_{3},\alpha_{4}\sigma_{4}}}^{irr}(i\nu,i\nu^{\prime})_{i\omega}=\frac{1}{T}[(\chi_{loc}^{0})_{i\omega}^{-1}-\chi_{loc}^{-1}].
\end{equation}
$\Gamma_{loc}^{irr}$ depends on three Matsubara frequencies ($i\nu$,
$i\nu^{\prime}$; $i\omega$), and both the spin ($\sigma_{1-4}$) and the orbital
($\alpha_{1-4}$) indices, which run over $3d$ states on the iron
atom. $T$ is the temperature.

Once the irreducible vertex $\Gamma_{loc}^{irr}$ is obtained, the
momentum dependent two-particle Green's function is constructed again
using the Bethe-Salpeter equation (Fig.~\ref{fig:BSE}) by replacing the
local polarization function $\chi_{loc}^{0}$ by the non-local one
$\chi_{\textbf{q},i\omega}^{0}$:
\begin{equation}
\chi_{{\alpha_{1}\sigma_{1},\alpha_{2}\sigma_{2}\atop
    \alpha_{3}\sigma_{3},\alpha_{4}\sigma_{4}}}(i\nu,i\nu^{\prime})_{\textbf{q},i\omega}=[(\chi^{0})_{\textbf{q},i\omega}^{-1}-T\cdot\Gamma_{loc}^{irr}]^{-1}.
\label{eq:BSE_imag}
\end{equation}
Finally, the dynamic
magnetic susceptibility $\chi(\textbf{q},i\omega)$ is obtained by
closing the two particle green's function with the magnetic moment
$\mu=\mu_B(\textbf{L}+2\textbf{S})$ vertex, and 
summing
over frequencies ($i\nu$,$i\nu^{\prime}$), orbitals ($\alpha_{1-4}$),
and spins ($\sigma_{1-4}$) on the four external legs
\begin{equation}
\chi(\textbf{q},i\omega)=T\sum_{i\nu,i\nu^{\prime}}\sum_{{\alpha_{1}\alpha_{2}\atop\alpha_{3}\alpha_{4}}}\sum_{{\sigma_{1}\sigma_{2}\atop
    \sigma_{3}\sigma_{4}}}\mu_{{\alpha_{1}\sigma_{1}\atop\alpha_{3}\sigma_{3}}}^{z}
\mu_{{\alpha_{2}\sigma_{2}\atop\alpha_{4}\sigma_{4}}}^{z} \;\chi_{{\alpha_{1}\sigma_{1},\alpha_{2}\sigma_{2}\atop
    \alpha_{3}\sigma_{3},\alpha_{4}\sigma_{4}}}(i\nu,i\nu^{\prime})_{\textbf{q},i\omega}
\label{eq:chi}
\end{equation}

The resulting dynamical magnetic susceptibility is obtained in
Matsubara frequency ($i\omega$) space and it needs to be analytically
continued to real frequencies ($\chi(\textbf{q},\omega)$). For the low
frequency region, on which we concentrate here, the vertex
$\Gamma_{loc}^{irr}$ is analytically continued by a quasiparticle-like
approximation.
We replace the frequency dependent
vertex with a constant, i.e.,
$\Gamma_{loc{\alpha_{1}\sigma_{1},\alpha_{2}\sigma_{2}\atop
    \alpha_{3}\sigma_{3},\alpha_{4}\sigma_{4}}}^{irr}(i\nu,i\nu^{\prime})_{i\omega}
\approx \bar{U}_{{\alpha_{1}\sigma_{1},\alpha_{2}\sigma_{2}\atop
    \alpha_{3}\sigma_{3},\alpha_{4}\sigma_{4}}}$,
and require
$\chi(q,i\omega=0)=\chi(q,\omega=0)$. 
This vertex $\bar{U}$ however retains important spin and orbital dependence.
%The details are given in the Supplementary Material.

%
\begin{figure}
\includegraphics[scale=0.42]{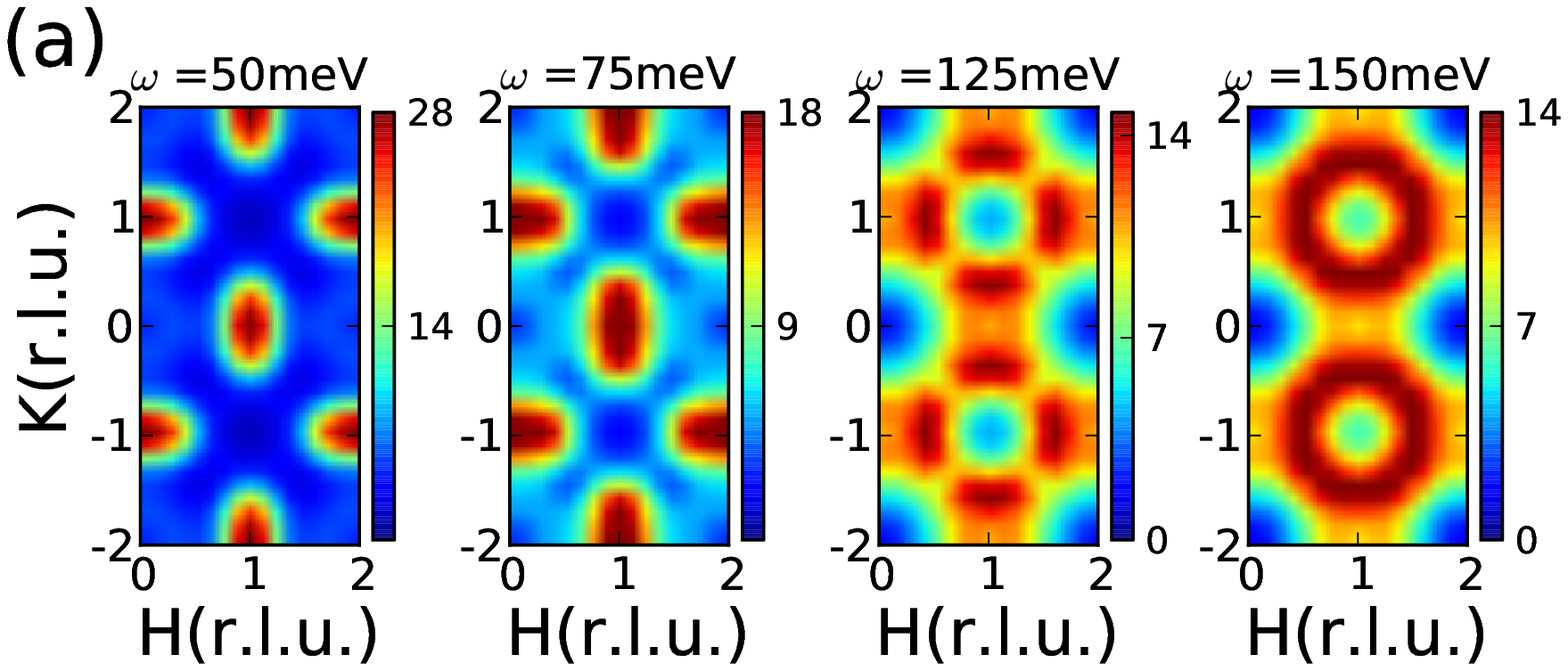}
\includegraphics[scale=0.42]{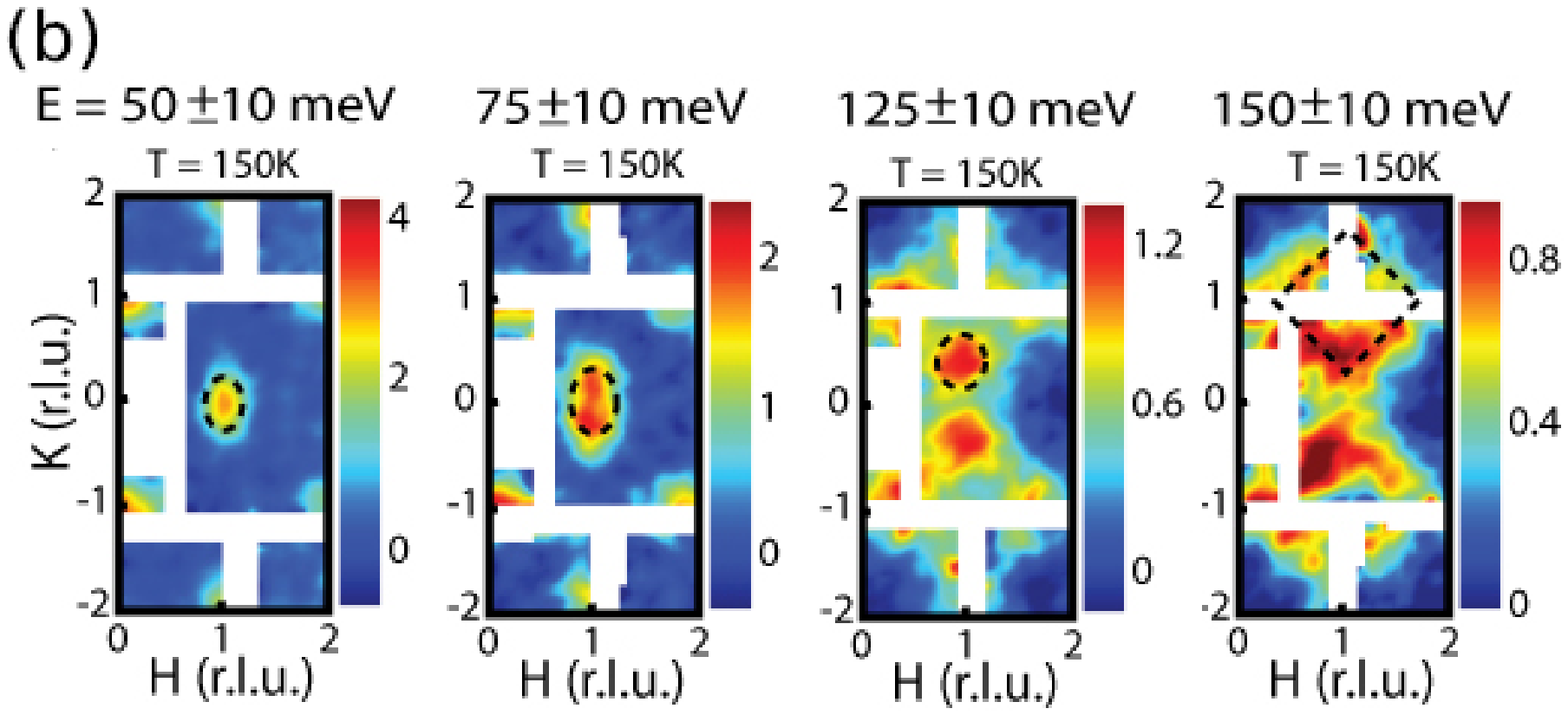}
\caption{(Color online) (a) The constant energy plot of the theoretical dynamical
  structure  factor, S($\textbf{q},\omega$)  
(=$\frac{\chi^{\prime\prime}(\textbf{q},\omega)}{1-e^{-\hbar\omega/k_{B}T}}$)
at different energies (50meV, 75meV, 125meV, and 150meV) in the
paramagnetic state (T=386$\,$K) of BaFe$_{2}$As$_{2}$ as a function of
momentum  $\textbf{q}=$(H,K,L).
%
%The interaction parameters are determined
%as $U$=5eV and $J$=0.7eV from the self-consistent GW
%calculation\cite{Kutepov:10}.
%
% The wave vector $q=$(H,K,L) is determined as ($q_{x}$a/$2\pi$,
% $q_{y}$b/$2\pi$, $q_{z}$c/$2\pi$).
L is here fixed at 1.
(b) The corresponding inelastic neutron scattering data from
Ref.~\onlinecite{Dai:10}.
\label{fig:Const-energy}}
\end{figure}

Fig.~\ref{fig:Const-energy}(a) shows the calculated constant energy
plot of the dynamical structure factor, S($\textbf{q},\omega$) in the
paramagnetic state of BaFe$_{2}$As$_{2}$. Our theoretical results are
calculated in the unfolded Brillouin zone of one Fe atom per unit
cell, because magnetic excitations are concentrated primarily on Fe
atoms, therefore folding, which occurs due to the two inequivalent arsenic
atoms in the unit cell, is not noticeable in magnetic
response~\cite{Keimer:10}.  For comparison we also reproduce in
Fig.~\ref{fig:Const-energy}(b) the INS experimental data from
Ref.~\onlinecite{Dai:10}. At low energy (around $\omega$=50meV), the
theoretical S($\textbf{q},\omega$) is strongly peaked at the ordering wave
vector (H,K,L)=$(1,0,1)$ and it forms a clear elliptical shapes elongated in K
direction. The
elongation of the ellipse increases with energy ($\omega$=75meV) and
around $\omega=$125meV the ellipse splits into two peaks, one peak
centered at $(1,0.4,1)$ and the other at $(1,-0.4,1)$.
%Very similar splitting is seen in experimental spectra in Fig.~\ref{fig:Const-energy}(b).
At even higher energy ($\omega\approx$150meV) the magnetic spectra
broadens and peaks from four equivalent wave vectors merge into a
circular shape centered at wave vector $(1,1,1)$.  At even higher
energy (230meV, not shown in the figure) the spectra broadens
further, and the peak becomes centered at the point $(1,1,1)$.
These trends are all in good quantitative agreement with INS data from
Fig.~\ref{fig:Const-energy}(b).

\begin{figure}
\includegraphics[bb=54bp 180bp 510bp 630bp,scale=0.38]{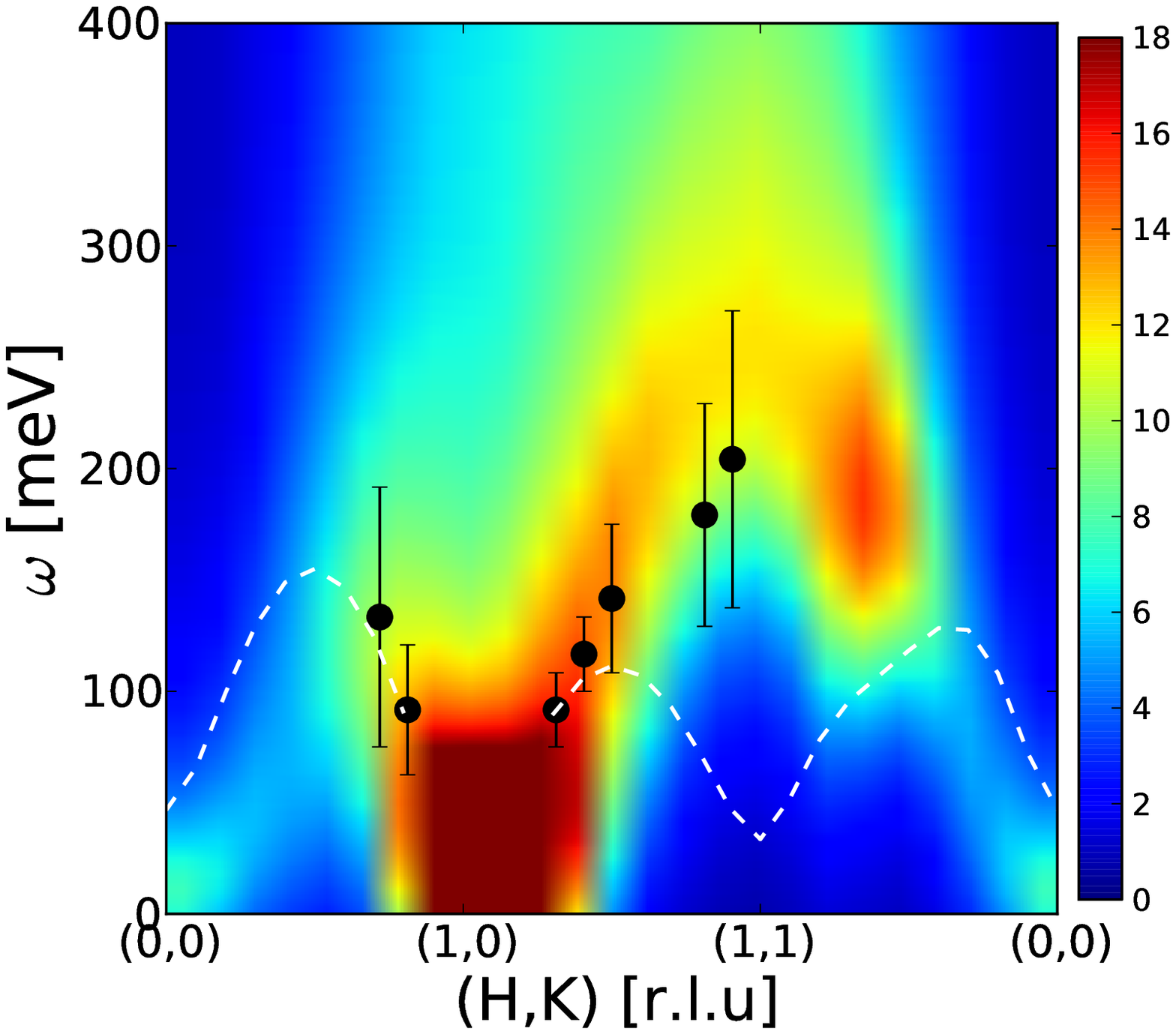}$\:\:\:$
\includegraphics[scale=0.08]{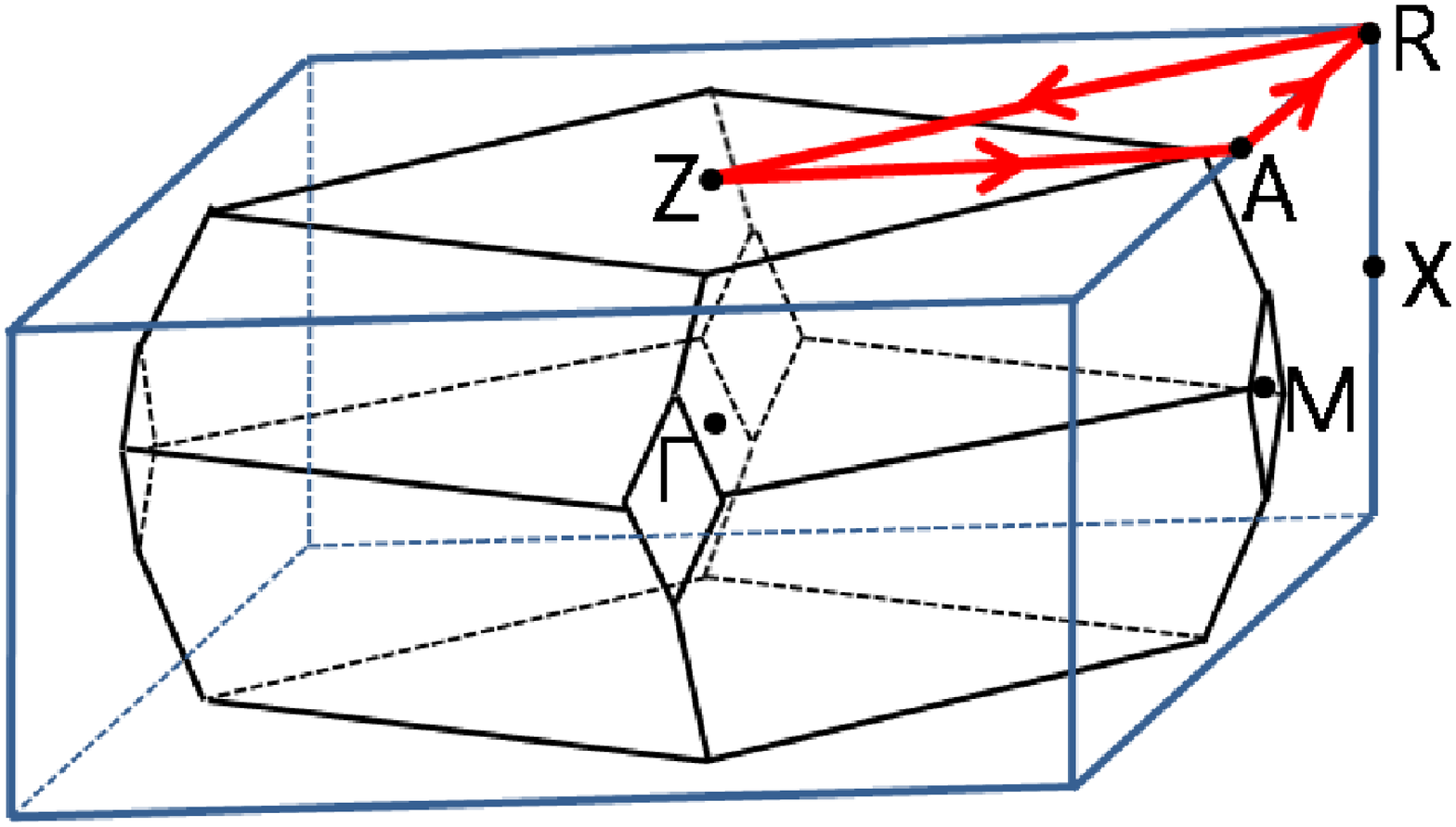}
\caption{(Color online) S($\textbf{q},\omega$) along the special path in Brillouin zone
  marked by red arrow in the inset on the right. The inset shows the
  body-centered tetragonal (black line) and the unfolded (blue line) Brillouin
  zone.
Black dots with error bars correspond to INS data from
Ref.~\onlinecite{Dai:10}.
The white dashed line shows the isotropic Heisenberg spin wave
dispersion.}
\label{fig:S_q_w}
\end{figure}

In Fig.~\ref{fig:S_q_w}, we display a contour plot of the  theoretical
S($\textbf{q},\omega$) as a function of frequency $\omega$ and
momentum $\textbf{q}$ along the special path in the unfolded Brillouin
zone, sketched by a red line in the right figure.  At low energies
($\omega<$80meV), S($\textbf{q},\omega$) is mostly concentrated in the
region near the ordering vector $(1,0,1)$. Consistent with the
elongation of the ellipse along the K direction in
Fig.~\ref{fig:Const-energy}, the low energy ($\omega<$80meV) bright
spot in Fig.~\ref{fig:S_q_w} is extended further towards $(1,1,1)$
direction but quite abruptly decreases in the $(0,0,1)$ direction.  The
magnetic spectra in the two directions $(1,0,1)\rightarrow (0,0,1)$ and
$(1,0,1)\rightarrow (1,1,1)$ are clearly different even at higher energy
$\omega>100$meV.  The peak position is moving to higher energy along
both paths, but it fades away very quickly along the first path, such
that the signal practically disappears at $(0.5,0,1)$. Along the second
path $(1,0,1)\rightarrow (1,1,1)$, there remains a well defined excitation
peak for which the energy is increasing, and at $(1,1,1)$ reaches the
maximum value of $\approx 230$meV. Continuing the path from
$(1,1,1)$ towards $(0,0,1)$ the peak energy decreases again and it fades
away around $(0.5,0.5,1)$. The  black dots
display INS data with errors bars from Ref.~\onlinecite{Dai:10}.  Notice a very good
agreement between theory and experiment.

The white dashed line in
Fig.~\ref{fig:S_q_w} represents the spin wave dispersion obtained for
the isotropic Heisenberg model using nearest neighbor $J_1$ and next
nearest neighbor $J_2$ exchange constants and performing the best fit
to INS data. This fit was performed in Ref.~\onlinecite{Dai:10}.  The
magnetic excitation spectra of an isotropic Heisenberg model show a
local minimum at the wave vector $\textbf{q}=(1,1,1)$, which is
inconsistent with our theory and with the experiment. To better fit the
experimental data with a Heisenberg-like model, very anisotropic
exchange constants need to be assumed~\cite{Dai:10}, which raised
speculations about possible existence of nematic phase well above the
structural transition of BaFe$_2$As$_2$. Since the DFT+DMFT results
can account for all the features of the measured magnetic spectra
without invoking any rotationally symmetry breaking  
the presence of nematicity in the
paramagnetic tetragonal state at high temperature  is  unlikely.
\begin{figure}
\includegraphics[scale=0.4]{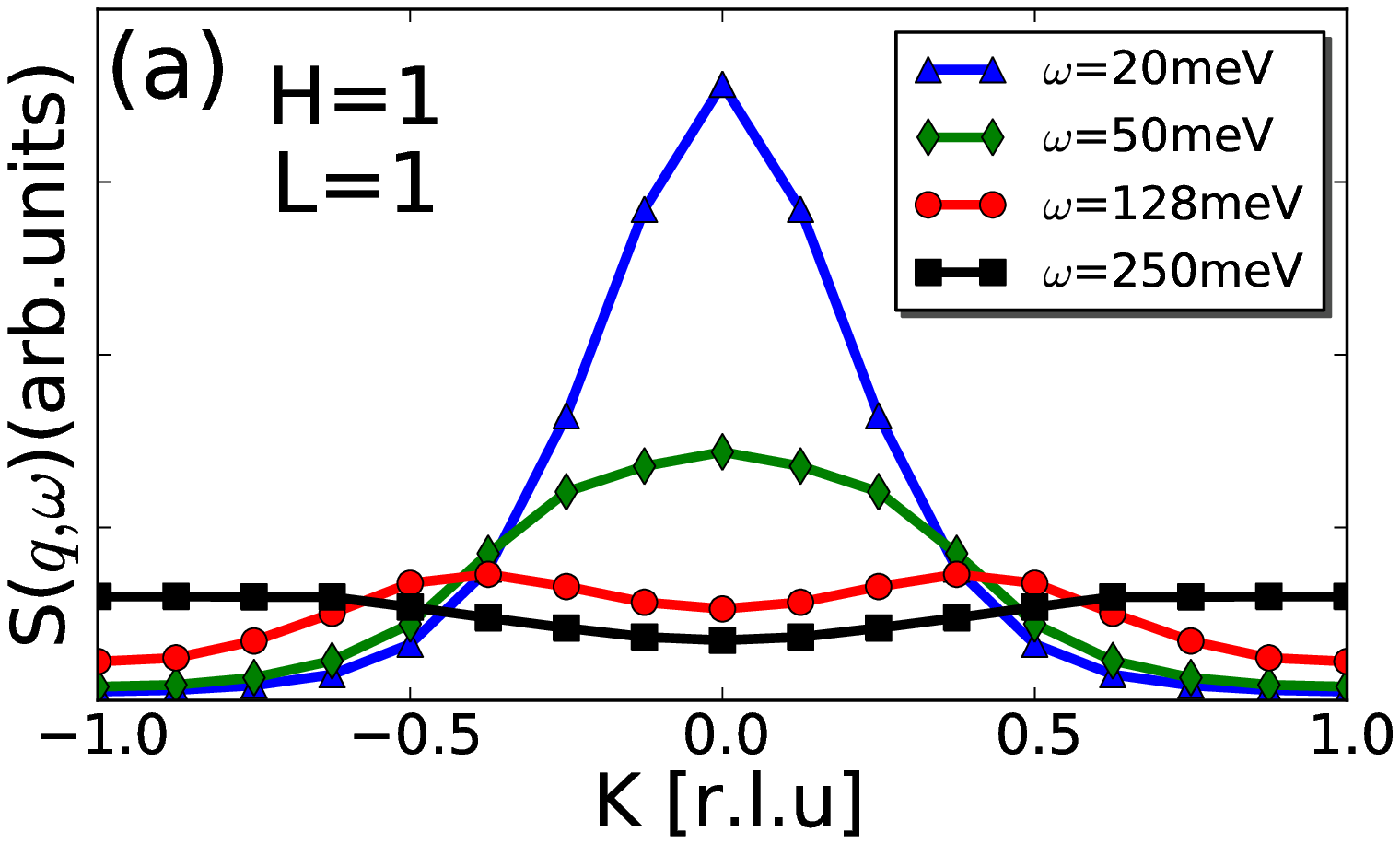}
\includegraphics[scale=0.4]{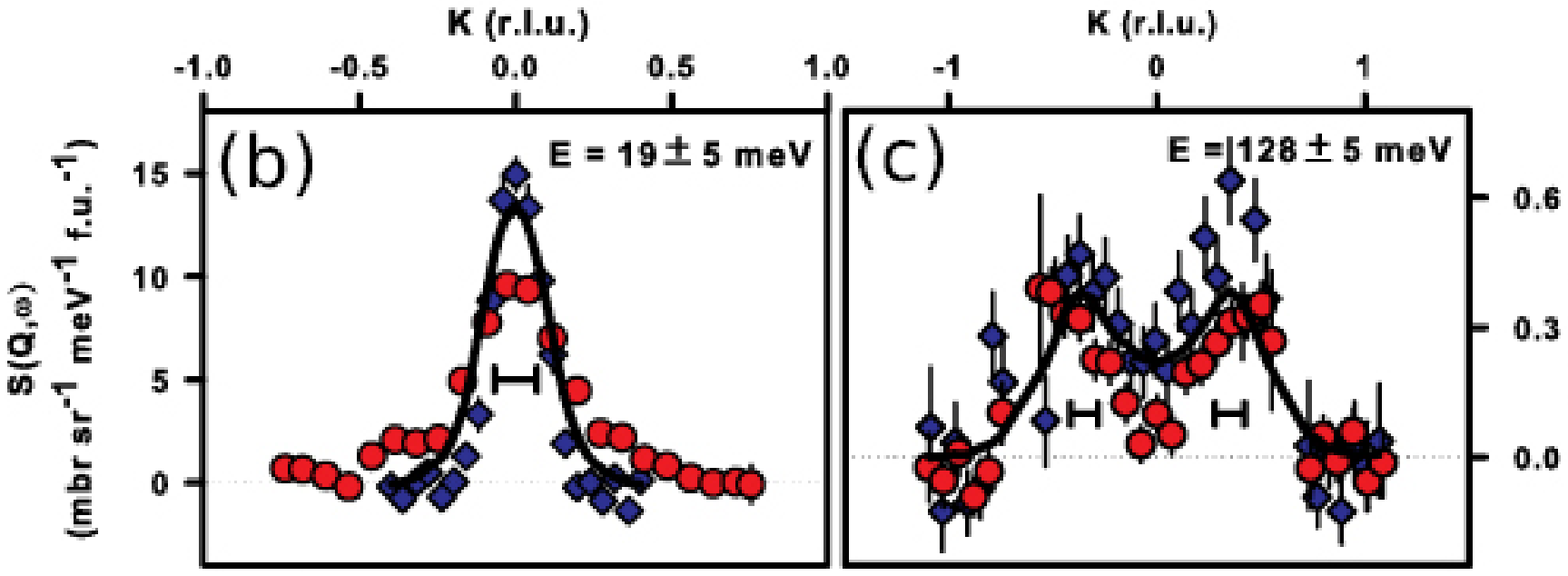}
\caption{(Color online) (a) The wave vector K dependence (H=1,L=1) of S($\textbf{q},\omega$) at
several frequencies. (b) The corresponding INS
data at $\omega$=19meV and (c) 128meV reproduced from
Ref.~\onlinecite{Dai:10}.
The red circles correspond to the paramagnetic state at $T$=150K and the
blue diamonds to the magnetic state at $T$=7K.
\label{fig:S_q}}
\end{figure}

In Fig.~\ref{fig:S_q}(a) we show constant frequency cuts in the K
direction (from $(1,-1,1)$ through $(1,0,0)$ to $(1,1,1)$) of
S($\textbf{q},\omega$) displayed in Fig.~\ref{fig:S_q_w}.  For comparison we
also show the corresponding INS measurements from
Ref.~\onlinecite{Dai:10} as red circles in Fig.~\ref{fig:S_q}(b) and (c).
%
% In order to examine further how the magnetic excitation peak changes
% as the energy $\omega$ increases, we plot S($q,\omega$) through
% the K direction fixing H=1, L=1 in the unfolded Brillouin zone at
% different energies (Fig.\ref{fig:S_q} top).
% 
At $\omega$=20meV, the spectrum has a sharp peak centered at the
ordering vector $(1,0,1)$.
%
%and agrees with the peak in the experiment
%(red circles in Fig.\ref{fig:S_q}(b)).
%
At $\omega$=50meV, the
spectrum still displays a peak at $(1,0,1)$ but the intensity is
significantly reduced.
With increasing frequency $\omega$, the peak position in S($\textbf{q},\omega$)
moves in the direction of $(1,1,1)$, and at 128meV peaks around
$(1,0.4,1)$. The shift of the peak is accompanied with substantial
reduction of intensity at ordering wave vector $(1,0,1)$.
At even higher energy of 250meV only a very weak peak remains,
and it is centered at the wave vector $(1,1,1)$.
The position of peaks as well as their frequency dependence
is in very good agreement with INS experiments of
Ref.~\onlinecite{Dai:10} displayed in Fig.~\ref{fig:S_q}(b) and (c).

\begin{figure}
\includegraphics[scale=0.39]{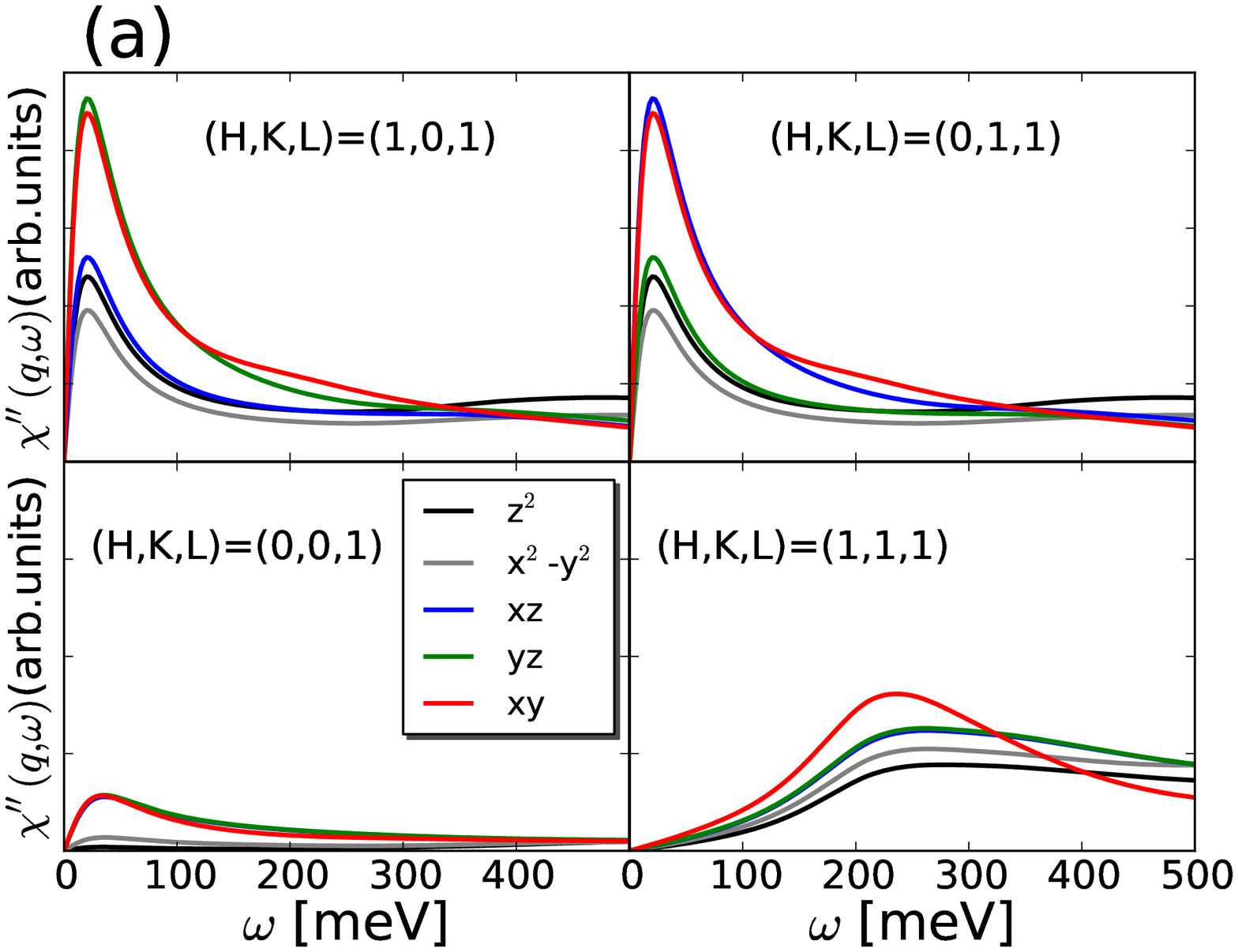}
\includegraphics[scale=0.4]{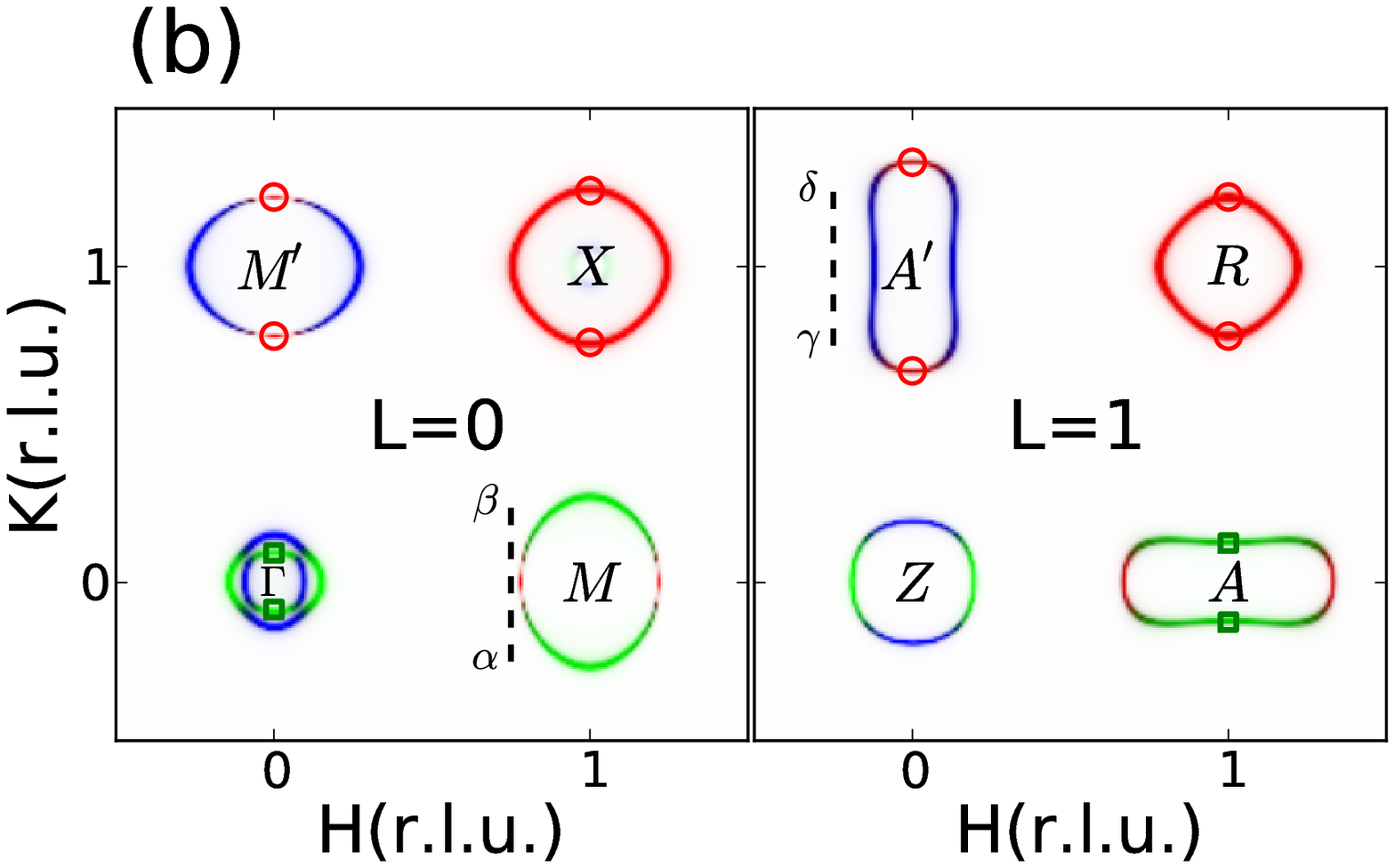}
\includegraphics[scale=0.35]{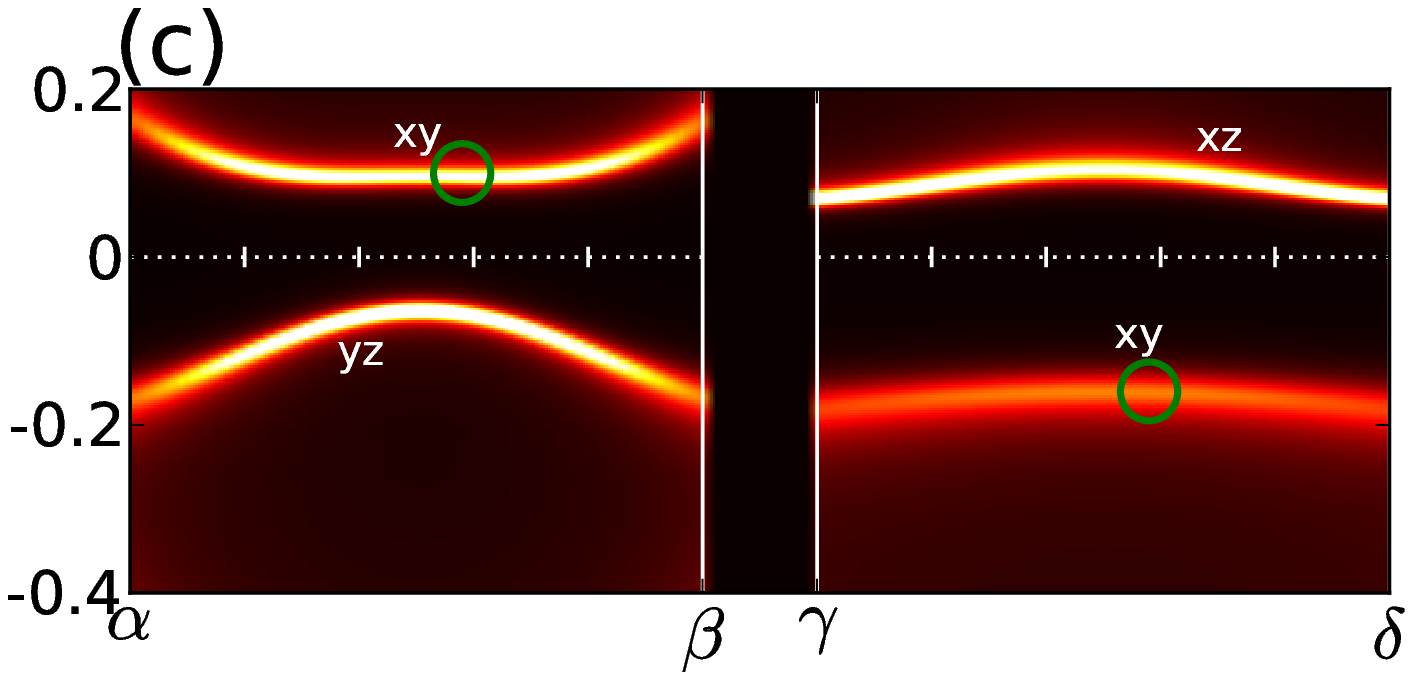}
\includegraphics[scale=0.4]{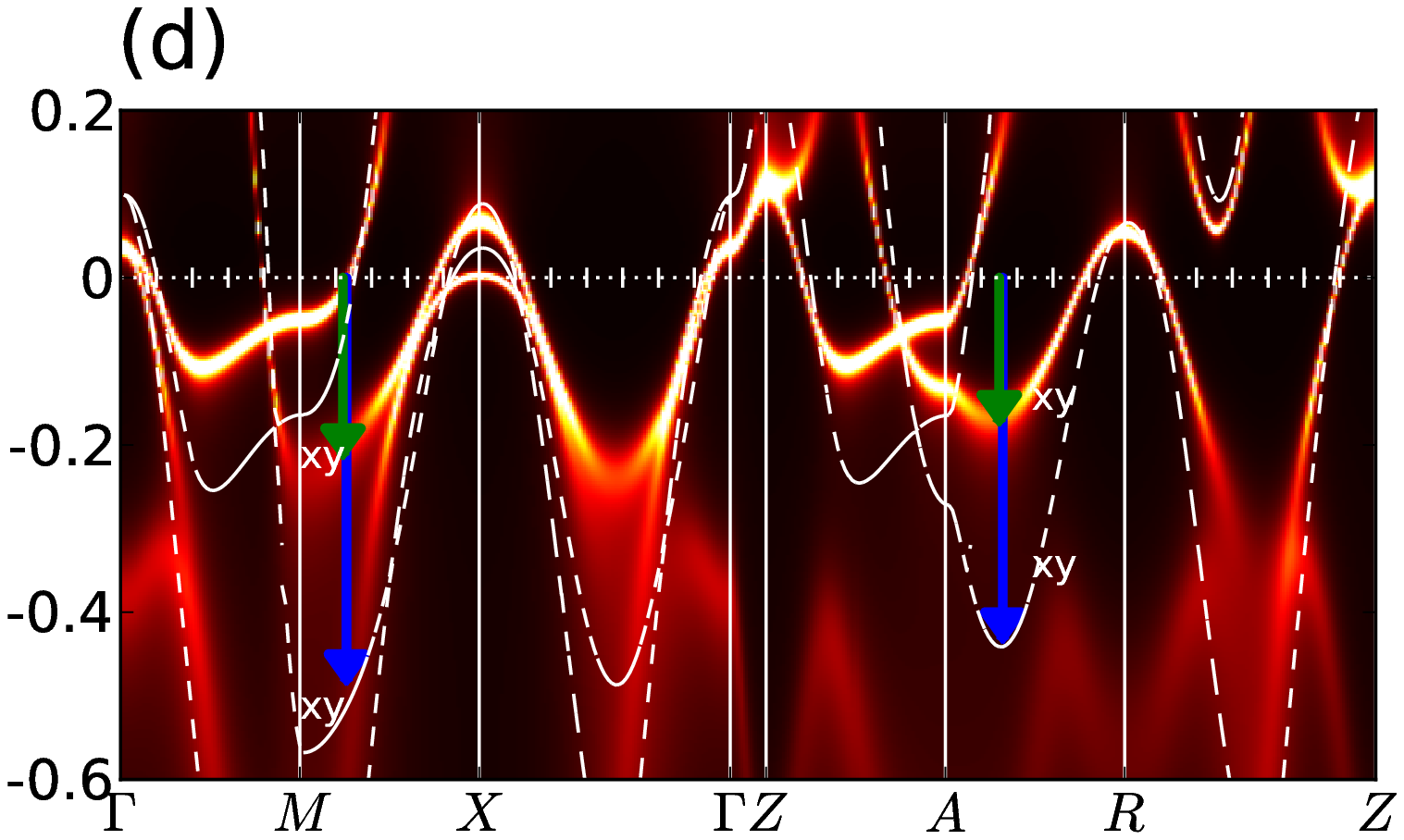}
\caption{(Color online) (a) The Fe $d$ orbital resolved dynamical magnetic susceptibility
at $T$=386K for distinct wave vectors $\textbf{q}$=$(1,0,1)$, $(0,1,1)$, $(0,0,1)$,
and $(1,1,1)$. Different colors correspond to different orbital
contributions.
(b) The Fermi surface in the $\Gamma$ and $Z$ plane at $T$=73K
colored by the orbital characters: $d_{xz}$(blue),
$d_{yz}$(green), and $d_{xy}$(red).
The small symbols mark the regions
in the Fermi surface, where nesting for the wave vector
$\textbf{q}$=$(1,0,1)$ is good.
%The green filled squares $\square$ (red filled
%triangles $\triangle,\nabla$) represent the $d_{yz}$ ($d_{xy}$) 
%contribution.
(c) The zoom-in of $A(\textbf{k},\omega)$ along the path marked by
black dashed line in Fig.~\ref{fig:chi_band_FS}(b). The green open circles 
indicate the two
relevant bands of $d_{xy}$ character which give rise to the peak in
magnetic excitation spectra near 230meV at $\textbf{q}$=$(1,1,1)$.
(d) $A(\textbf{k},\omega)$ computed in
one Fe atom per unit cell at $T$=73K. The green arrows mark the same bands
which give rise to 230meV peak.
The DFT bands are overlayed
by white dashed lines. The blue arrows mark corresponding DFT bands of
$d_{xy}$ character.}
\label{fig:chi_band_FS}
\end{figure}

Fig.~\ref{fig:chi_band_FS}(a) resolves the dynamical magnetic
susceptibility $\chi$ of Eq.~\ref{eq:chi} in the orbital space
$\chi_{\alpha}=T\sum_{i\nu,i\nu^{\prime}}\sum_{\beta}\sum_{{\sigma_{1}\sigma_{2}\atop
    \sigma_{3}\sigma_{4}}}\mu_{\sigma_{1}\sigma_{3}}^{z}
\mu_{\sigma_{2}\sigma_{4}}^{z} \;\chi_{{\alpha\sigma_{1},\beta\sigma_{2}\atop
    \alpha\sigma_{3},\beta\sigma_{4}}}(i\nu,i\nu^{\prime})$
such that $\chi=\sum_{\alpha}\chi_\alpha$.
At the magnetic ordering vector $\textbf{q}$=$(1,0,1)$, 
$\chi_{\alpha}^{\prime\prime}$ increases sharply with frequency
near $\omega=0$ for all orbitals and is strongly suppressed above
100meV reaching the maximum around 20meV.
At this wave vector, the dominant contributions at low energy come
from the $d_{xy}$ and the $d_{yz}$ orbitals.
The magnetic susceptibility at $\textbf{q}$=$(0,1,1)$ in
Fig.~\ref{fig:chi_band_FS}(a) shows the same trend as orbitally resolved
spectra at $\textbf{q}$=$(1,0,1)$, except that $d_{xz}$ and $d_{yz}$ switch their
roles due to the $C_{4}$ symmetry of the Fe square lattice.

These dominant orbital contributions to $\chi$ are also reasonably
captured in the polarization bubble $\chi^0$ (not shown here), hence
these excitations could be understood in terms of the Fermi surface nesting.  The
orbital resolved Fermi surface is displayed in
Fig.~\ref{fig:chi_band_FS}(b) at both the $\Gamma$-plane and the
$Z$-plane. Most of
the weight in $\chi^0$ comes from the diagonal terms, i.e.,
$\chi^0_{\alpha,\alpha}$, hence the Fermi surfaces with the same color
in Fig.~\ref{fig:chi_band_FS}(b) but separated by the wave vector
$(1,0,1)$ give dominant contribution.
% 
% The orbital resolved Fermi
% surface (Fig. \ref{fig:chi_band_FS}(b)) shows that the intra-orbital
% transitions between $d_{xy}$ (red) orbitals and $d_{yz}$ (green)
% orbitals are dominant as the particle-hole excitation for the nesting
% vector (1,0,1).
%
The intra-orbital $d_{yz}$ low energy spectra comes mostly from the
transitions between the green parts of the hole pocket at $\Gamma$ and
the green parts of the electron pocket at $A$, marked with green squares
($\square$) in Fig.~\ref{fig:chi_band_FS}(b).
Since the electron pocket at $A$ is elongated in H direction, the
nesting condition occurs mostly in the perpendicular K direction,
hence the elliptical excitations at low energy in
Fig.~\ref{fig:Const-energy} are elongated in K but not in H direction.
The intra-orbital $d_{xy}$ transitions are pronounced between the
electron pocket at $M'$ and the hole pocket at $R$, 
as well as between the electron pocket at $A'$ and the hole
pocket at $X$ (marked with red $\bigcirc$).
This large spin response at
$(1,0,1)$ gives rise to the low energy peak in Fig.~\ref{fig:S_q_w}.

% The intra- and inter-orbital two-particle interactions enhance magnetic
% excitation spectra resulting in the collective excitation between
% quasi-particles.

%The polarization bubble $\chi^0_{\alpha}$ is very small for $eg$
%orbitals, since $eg$ orbitals have very small one particle density of
%states at the Fermi level, however, the magnetic susceptibilty
%$\chi_{\alpha}$ is significant also for $eg$ orbitals. This is due to
%the large off-diagonal terms in the vertex $\Gamma$, which is
%inherited from the off-diagonal nature of the Hund's coupling
%interaction $J$.

% At the wave vector (1,0,1) in Fig. \ref{fig:chi_band_FS}(a), $d_{xz}$,
% $d_{x^{2}-y^{2}}$, and $d_{z^{2}}$ orbitals, which are strongly
% suppressed in the one-particle excitation ($\chi^{0}$), show
% non-negligible contributions to the full magnetic excitation spectra
% ($\chi$) at low energies. Specially, the inter-orbital two-particle
% interaction corresponding to the Hund's coupling $J$, plays an
% important role to enhance the full magnetic excitation spectra for all
% orbitals.

We note that the particle-hole response, encoded in polarization
bubble $\chi^0$, is especially large when nesting occurs between an
electron pockets and a hole pocket, because the nesting condition
extends to the finite frequency, and is not cut-off by the Fermi
functions.

The low energy magnetic excitations at wave vectors $\textbf{q}=(0,0,1)$ and
$\textbf{q}=(1,1,1)$ can come only from electron-electron or hole-hole
transitions, hence both responses are quite small, as seen in
Fig.~\ref{fig:chi_band_FS}(a).  While the magnetic response at
$\textbf{q}=(0,0,1)$ is small but finite, the spin response at $\textbf{q}=(1,1,1)$ is
almost gapped. This is because the hole-hole transitions from $\Gamma$
to $R$ or electron-electron transitions from $M$ to $A'$ do not
involve any intra-orbital transitions, and hence are even smaller than
transitions at the wave vector $(0,0,1)$.

At finite energy transfer, the spin excitations come from electronic
states away from the Fermi energy, and can not be easily identified in
the Fermi surface plot. Hence it is more intriguing to find the
dominant contribution to the peak at $\omega\approx 230$meV and
$\textbf{q}=(1,1,1)$. This peak gives rise to the 230meV excitations at $(1,1,1)$
in Fig.~\ref{fig:S_q_w}.
A large contribution to this finite frequency excitation comes from a
region near the two electron pockets at $M$ and $A'$ marked with black
dashed line in Fig.~\ref{fig:chi_band_FS}(b). We display in
Fig.~\ref{fig:chi_band_FS}(c) the one electron spectral function
across these dashed lines in the Brillouin zone to show an important
particle hole transition
from the electrons above Fermi level at the $M$
point and the flat band at -200meV around the $A'$ point, both of $d_{xy}$
character.
We note that due to large off diagonal terms in the two
particle vertex $\Gamma$, all orbital contributions to $\chi$ develop
a peak at the same energy, although only $d_{xy}$ orbital displays a
pronounced peak in $\chi^0$.

Fig.~\ref{fig:chi_band_FS}(d) displays the one electron spectral
function in a path through the Brillouin zone, corresponding to one Fe
atom per unit cell.  Within DFT+DMFT the quasi-particle bands are
renormalized by a factor of 2-3 compared to the corresponding DFT
bands (white dashed lines). The green arrow marks the $d_{xy}$ band
which contributes to the peak in $S(\textbf{q},\omega)$ near 230meV
and $\textbf{q}=(1,1,1)$. In DFT calculation, this $d_{xy}$ intra-orbital
transition is also present, but occurs at much higher energy of the
order of 400-600meV, marked by blue arrows.  The over-estimation of
the peak energy at $\textbf{q}$=$(1,1,1)$ was reported in LSDA calculation of
Ref.~\onlinecite{Ke:11}.

% The spin excitation
% energy at $q$=(1,1,1) in FeTe shows a smaller energy
% scale\cite{Lumsden:10} possibly due to the stronger renormalization of
% the $d_{xy}$ band in FeTe than BaFe$_{2}$As$_{2}$.

In this Letter, we have extended the DFT+DMFT methodology to compute
the two particle responses in a realistic multi-orbital DFT+DMFT
setting. With the same parameters which were used to successfully
describe the optical spectra and the magnetic moments of this
material~\cite{Yin:11}, we obtained a coherent description of the
experimental neutron scattering results. Our theory ties the magnetic
response to the fermioloy of the model, and quantifies the departure
from both purely itinerant and localized pictures.

$\textit{Acknowlegements:}$
We are grateful to Pencheng Dai for discussions of his experimental results.
This research was supported in part by the National Science Foundation through TeraGrid resources provided by Ranger (TACC) under grant number TG-DMR100048.

\bibliography{Pnictide}

\end{document}